\documentstyle[twoside,fleqn,espcrc2]{article}

\newcommand{\be}{\begin{equation}}
\newcommand{\ee}{\end{equation}}
\newcommand{\bea}{\begin{eqnarray}}
\newcommand{\eea}{\end{eqnarray}}

\begin{document}

\title{Recent results on self-dual configurations on the 
torus\thanks{presented by A. Gonz\'alez-Arroyo}}

\author{M. Garc\'{\i}a P\'erez ${}^{\rm a}{}$,  A. Gonz\'alez-Arroyo ${}^{\rm a}{}$,
A. Montero ${}^{\rm a}{}$, C. Pena
 \address{ Departamento de F\'{\i}sica Te\'orica C-XI,
        Universidad Aut\'onoma de Madrid, Madrid 28049, Spain.}
and P.~van~Baal\address{ Instituut-Lorentz for Theoretical Physics, 
University of Leiden, PO Box 9506, NL-2300 RA Leiden, The Netherlands.}}

\begin{abstract}
We review the recent  progress on our understanding of self-dual 
SU(N) Yang-Mills configurations on the torus. 
\end{abstract}

\maketitle

\section{Motivation}
The purpose of this paper is to briefly review the progress made during  the
last year in the investigation of  $SU(N)$ self-dual Yang-Mills (YM) configurations 
on the 4-dimensional torus ($T^4$). It is well-known that there is a procedure 
(the ADHM construction~\cite{ADHM}) to construct all self-dual 
configurations on the sphere. Conversely, for the torus case not even the
 simplest fully non-abelian solution is known analytically (abelian-like 
solutions are known for some torus sizes). Constructing them provides a 
challenge for both physicists and mathematicians. 

From the Physics point of view, compactification on the sphere amounts to the 
condition of finite action implying              that  the configuration 
approaches a pure gauge at infinity. This       is not a
 physically necessary condition even at a  purely classical level.
Since the action is an extensive quantity it 
makes more sense to demand finiteness of the action density. 
 In the Confinement regime,         it is reasonable 
to expect that typical configurations are different from the classical vacuum 
almost everywhere. On the other hand, configurations on the torus can be 
looked at as periodic configurations on $R^4$ having  finite action density. 
Although  periodicity is also an unphysical feature for the dominant Yang-Mills
configurations to have, knowledge of these configurations might help us to 
understand better some  structures which might be present in the vacuum. 

Besides, there are some cases in which one deals with periodicity in some 
of the variables. This can be seen as formulations on the non-compact 
manifolds $T^n\times R^{4-n}$. The case $n=1$ occurs naturally when studying 
field theory at finite temperature. The case $n=3$ has been studied in 
relation with the Hamiltonian formulation for YM fields on the torus. 
Recently~\cite{vortex},  it has been seen how the $n=2$ case is relevant 
in constructing $SU(N)$ YM vortex-like configurations in $R^4$.  
It is an interesting question to investigate how the $T^4$ configurations
are related to these non-compact manifold ones.

\section{Introduction}
Yang-Mills fields on the torus are classified by the topological charge $Q$ 
and by the twist sectors (see Ref.~\cite{review} for an introduction to the
 subject and a review of older results). The latter are a discrete number of 
sectors labelled by two 3-vectors of integers modulo $N$ ($\vec{k}$ and 
$\vec{m}$). The possible values of the topological charge are restricted by 
twist:
\be
Q= -\frac{\vec{k}\cdot \vec{m}}{N} + {\rm integer}\quad .
\ee
Orthogonal twists are those for which $\vec{k}\cdot \vec{m}= 0\, \bmod N$
(the rest are called non-orthogonal). 
Only in this case the topological charge is an integer (and the action a 
multiple of $8 \pi^2$).  

Self-dual solutions, if existing, form a manifold whose dimensionality (up to 
gauge transformations) is given by $4QN$.  Four of these modes correspond to 
space-time translations of the solution. Existence and non-existence has been 
proved in some cases. Particularly relevant is the  non-existence  
of $Q=1$ self-dual configurations on the torus without twist~\cite{PBPvB}
 ($\vec{k}=\vec{m}=0\, \bmod N$).

Apart from the afore-mentioned quasi-abelian solutions, most of what is 
known about these solutions comes from numerical studies on the lattice. 
There is, however, an important duality (involutive)  transformation --the Nahm 
transform~\cite{nahm}-- which  maps $SU(N)$ self-dual configurations on 
the torus with topological charge $Q$, onto  $SU(Q)$ self-dual configurations 
on the dual torus with topological charge $N$. Unfortunately, besides its role
in the proof of Ref.~\cite{PBPvB}, 
little use has been made of this property to increase our knowledge on these 
configurations. Part of the progress we report, has to do with fixing this
 state of affairs. 

The study of self-dual configurations depends on the group rank, twist, topological
 charge and torus size. For simplicity the study has focused on $SU(2)$
and low values of the topological charge ($Q=\frac{1}{2},1$). This forces 
non-zero twist. For obvious symmetry reasons the study has centered on 
hypercubical tori of size $l_s^3\times l_t$. For large and small aspect 
ratios $l_s/l_t$ one can study and exploit the connection with $R^3\times S_1$
and $T^3\times R$ configurations.  Now we will summarise what was known 
of these configurations before.

$\mathbf{l_t/l_s \gg 1}\,$\cite{numerical}:\
  For $Q=\frac{1}{2}$ (non-orthogonal twist) the solution is 
unique up to 
translations. The configuration is exponentially localised in time and as  $l_t/l_s$ 
goes to infinity  approaches a configuration known as the twisted instanton. 
For $Q=1$ the description of the resulting configuration depends on twist. For 
$\vec{m}\ne 0$ and $\vec{k} = 0 \bmod N=2$ (Space-like twist) the configuration is generically 
described by 2  twisted instantons separated in time. The 2 space-time locations describe 
the 8 parameters of the moduli space. For $Q=1$, $\vec{k}\ne 0$ and 
$\vec{m} = 0 \bmod 2$ (Time-like twist) one gets a family of (exponentially localised in
 time)  configurations which as $l_t/l_s$ goes to infinity  can be parametrised by the 
holonomies: the spatial Polyakov loops at infinite time. 

 $\mathbf{l_s/l_t \gg 1}\,$:\ In this case the $R^3\times S_1$ configurations
with $Q=1$ are known analytically~\cite{calorons}. Generically, they are given by 2 lumps 
which for large separations are simply BPS monopoles of various masses, fixed 
by the holonomy of the configuration (the time-like Polyakov loop at spatial 
infinity). 
We recently studied, by lattice methods, the large $l_s/l_t$ configurations 
on the torus~\cite{Calo}. These configurations neatly approach the analytic 
calorons with some restrictions on the moduli space. For $Q=\frac{1}{2}$ a 
single fixed mass periodic BPS monopole is obtained. For $Q=1$ and time-like 
twist one obtains in each cell a couple of equal mass constituent monopoles 
with arbitrary locations. Conversely for $Q=1$ and space-like twist 
one gets a couple of variable mass monopoles with fixed relative positions.

\section{New results}
In the last year there have been several developments  
 which we will now list:
\begin{itemize}
\item A numerical method, based on lattice gauge theories,
 has been developed to implement the NT
 numerically~\cite{nahmlat}. It allows to obtain the NT of lattice configurations
 approximating the self-dual continuum ones. The method has proved quite 
 precise and stable.
\item It has been shown how to extend the definition of the Nahm transform 
to the non-zero twist case~\cite{NahmTBC}. The extension preserves the main 
properties of 
the original NT. The transform of an   $SU(N)$ self-dual configuration on 
the torus with topological charge $Q$, is now an   $SU(Q N_0)$ self-dual
 configuration on some Nahm-dual  torus with topological charge $N/N_0$.
 The integer $N_0$ depends on twist and equals $1$ for zero twist.
\item An analysis of the NT of $Q=1$ self-dual configurations on $T^3\times R$
 for twisted boundary conditions in time has been performed\cite{nahm1}. 
Although, the original self-dual configuration is not known, its NT is known 
to be  an abelian field in $T^3$, which is self-dual everywhere except at 
certain pointlike singularities. These singularities act as dyonic sources 
and their location is
determined by the holonomies (spatial Polyakov loop at infinite time) of the 
original $T^3\times R$ configuration. With this information one is able to construct  this Nahm-dual abelian field everywhere.  
\item The numerical investigation of the periodic calorons mentioned before~\cite{Calo}. 
\end{itemize}

Equipped with this new information and techniques we recently 
studied~\cite{dualities} the action of the NT on the torus configurations 
under concern. One finds that configurations with large and small $l_s/l_t$ 
are mapped onto each other by the NT. Furthermore, for $Q\!=\!1$ the time-like or 
space-like nature of the twist is preserved by the transformation.
The results can be summarised as follows:
\begin{itemize}
\item The $Q\!=\frac{1}{2}$ twisted instanton maps onto the $Q\!=\frac{1}{2}$
periodic caloron. This can be tested by the numerical NT and shows remarkable 
precision (see fig.~1 of the second ref. in~\cite{nahmlat}).
\item For $Q\!=\!1$ and time-like twist we see that the approximate holonomy 
of the $l_t/l_s \gg 1$ periodic instanton, maps onto the relative position of 
two equal mass constituent monopoles. Taking the limit $l_t\rightarrow \infty$
($l_s$ fixed) we see that the dyonic singularities of the  NT of $T^3\times R$ 
confs, are nothing but BPS monopoles with the non-abelian cores (of size 
$1/l_t$) shrunk to 0.
\item For $Q=1$ and space-like twist, the holonomy of the large $l_t/l_s$ 
configuration is fixed, which explains the fixed relative position of the 
constituent monopoles of the Nahm transform. Furthermore, the time distance 
between the 2 twisted instantons of this configuration  maps onto the 
holonomy (and hence the monopole masses) of the periodic caloron configuration.
This is clearly depicted in Fig.~1.
\end{itemize} 
For details the reader is referred to \cite{dualities}

In summary, all the known information is nicely linked 
non-trivially together by the  NT. A general pattern  mapping  
 approximate holonomies to lump positions emerges from our study.

\begin{figure}[htb]
\vspace{1.0cm}
\includegraphics{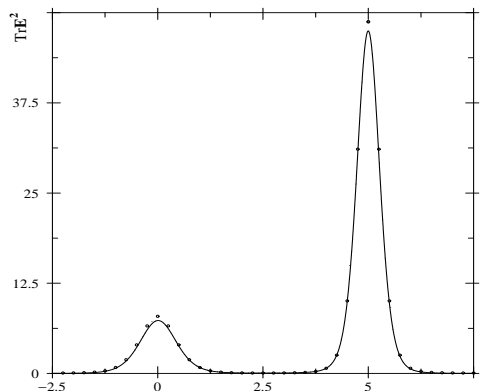}
\vspace{3.3cm}
\caption{ Comparison of the NT of an $l_t/l_s \gg 1$ configuration with
the analytic  caloron profile for the predicted unequal mass constituent monopoles
 (no free parameters). 
}
\end{figure}

\end{document}